# The Superspinorial Field Theory in Riemannian Coordinates*

Yaroslav Derbenev

derbenev@jlab.org

*Thomas Jefferson National Accelerator Facility, Newport News, VA 23606, USA*

Abstract

The *Superspinorial Dual-covariant Field Theory* (SSFT) developed in papers [1, 2] is treated in terms of *Riemannian coordinates* (RC) [7, 8] in space of the *N* dimensions *unified manifold* (UM). Metric tensor of UM (*grand metric*, GM) is built on the *split metric* matrices (SM) [1] which are a proportion of the *Cartan's affinors* (an extended analog of *Dirac's matrices*) of his *Theory of Spinors* [3] as explicated in [2]. Transition to RC based on consideration of *geodesics* is described. A principal property of an *orthogonal* RC frame (ORC) utilized in the present paper is *constancy* of the *rotation matrix A* of the *Riemannian space* of UM, while transformation matrix *B* of the *dual superspinorial state vector* field (DSV) varies together with Cartan's affinors according to the *dynamical law* of SSFT derived in [2]. The spinorial genesis of notion of the *orthogonality* as aspect of irreducible SSFT is pointed out in the present paper. The main outcome of resorting to an orthogonal RC frame (ORC) is explication of the *conformal dynamic invariance* of metric i.e. constancy of ratio between components of GM in ORC frame. This leads to possibility of reduction of algebraic equations of SSFT for the SM [2] to equation for GM *determinant* where GM *signature* plays a role of *N* discrete parameters non-specified in advance.

## 1. Introduction: a concise exposition of paper [2]

Here we will briefly reproduce contents of paper [2] complementing the exposition with a few comments.

### 1.1. Generally covariant Superdimensional FT

*Basic principles*

A Superdimensional dual-covariant field theory (SFT) has been presented in paper [1]. It is constituent based on a system of the *irreducibility* demands to a unified dynamic law for an ensemble of basic objects *X* which includes a multicomponent *dual state vector* (DSV) field $\Psi^\alpha \equiv \Psi$; $\Phi_\alpha \equiv \Phi$, *split metric* (SM) matrices (on Greek indices) $\Lambda^{\alpha k}_\beta \equiv \Lambda^k \equiv \Lambda$ and *unified gauge field* (UGF) matrices (shortly *gauge*) $\mathcal{A}^\alpha_{\beta k} \equiv \mathcal{A}_k \equiv \mathcal{A}$ as functions of *N* variables $\breve{\psi}^k$ of *unified manifold* (UM) ($k = 1, 2, \ldots N$; $\alpha, \beta = 1, 2, \ldots \mu$). *Differential Law* (DL) of SFT has been derived based on the *Extreme Action* principle (EAP):

$$\delta \int \mathcal{L}(X, \partial X) d\Omega = 0; \qquad d\Omega \equiv d\check{\psi}^1 d\check{\psi}^2 \ldots d\check{\psi}^N; \qquad (1.1)$$

where $\mathcal{L}$ is *Lagrangian form* structured on basic objects $X$ and their derivatives $\partial X$. Posing, as usual, variations of basic objects $\delta X = 0$ at a (arbitrary) closed surface limiting volume of integration, EAP generally results in *Euler-Lagrange* (EL) *equations* for system of basic objects:

$$\partial_k \frac{\partial \mathcal{L}}{\partial(\partial_k X^\alpha)} = \frac{\partial \mathcal{L}}{\partial X^\alpha}; \qquad \partial_k \equiv \frac{\partial}{\partial \check{\psi}^k}. \qquad (1.2)$$

Lagrangian $\mathcal{L}$ is product of *scalar Lagrangian* L and *weigh factor* $\sqrt{\Lambda}$:

$$\mathcal{L} \equiv \mathrm{L}\sqrt{\Lambda}; \qquad \Lambda \equiv |det\Lambda_{kl}| \qquad (1.3)$$

where $\Lambda_{kl}$ is metric tensor of UM, so $\sqrt{\Lambda} d\Omega$ is *invariant differential volume*. Lagrangian is composed based on a system of principles consolidated by demand *of irreducibility* of a *differential law* (DL) of SFT (*diffeohomomorphism, duality, reality, uniformity, homogeneity, differential irreducibility, covariance, existence of a conservative current, scaling invariance* and *mini-max principle*):

$$\mathrm{L} \Rightarrow \mathbb{L}; \qquad \mathbb{L} \equiv \mathbb{M} + \mathbb{G}; \qquad \mathbb{M} \equiv \mathbb{N} + \mathbb{D}; \qquad (1.4)$$

$$\mathbb{N} \equiv \Phi_\alpha \Psi^\alpha; \qquad \mathbb{D} \equiv \Lambda_\alpha^{\beta k} \mathfrak{D}_{\beta k}^\alpha; \qquad \mathfrak{D}_{\beta k}^\alpha \equiv \frac{1}{2}(\Phi_\beta \mathfrak{D}_k \Psi^\alpha - \Psi^\alpha \mathfrak{D}_k \Phi_\beta) \equiv \mathfrak{D}_k; \qquad (1.5)$$

$$\mathfrak{D}_k \Psi^\beta \equiv \partial_k \Psi^\beta + \mathcal{A}_{\gamma k}^\beta \Psi^\gamma \equiv \mathfrak{D}_k \boldsymbol{\Psi}; \qquad \mathfrak{D}_k \Phi_\beta \equiv \partial_k \Phi_\beta - \mathcal{A}_{\beta k}^\gamma \Phi_\gamma \equiv \mathfrak{D}_k \boldsymbol{\Phi} \qquad (1.6)$$

$$\mathbb{G} \equiv \frac{1}{4} \Lambda^{kl} \Lambda^{mn} \mathbb{G}_{km;ln}; \qquad \mathbb{G}_{km;ln} \equiv Tr(\mathfrak{R}_{km} \mathfrak{R}_{ln}); \qquad (1.7)$$

$$\mathfrak{R}_{kl} \equiv \mathfrak{R}_{\beta kl}^\alpha \equiv \partial_k \mathcal{A}_l - \partial_l \mathcal{A}_k + [\mathcal{A}_k, \mathcal{A}_l]; \qquad (1.8)$$

$$\Lambda^{kl} \equiv \frac{1}{\mu} Tr(\boldsymbol{\Lambda}^k \boldsymbol{\Lambda}^l) = \frac{1}{\mu} \Lambda_\beta^{\alpha k} \Lambda_\alpha^{\beta l}; \qquad \Lambda^{km} \Lambda_{lm} = \Delta_l^k = \begin{cases} 0; & l \neq k \\ 1; & l = k \end{cases} \qquad (1.9)$$

The Roman and Greek indices do not interfere, since they are associated with transformations in two different spaces, *unified manifold* (UM, variables $\check{\psi}^k$) and *matter function* (MF, variables $\varphi^\alpha$) with matrices $A$ and $B$, respectively [1]. Forms (1.6) are *covariant derivatives* of DSV. Form $\mathfrak{D}_{\beta k}^\alpha$ is named *matter matrices* (MM), and form $\mathfrak{R}_{\beta kl}^\alpha$ *hybrid curvature form*, HCF (symbol [ , ] denotes commutator of two matrices). HCF is uniquely recognized as *covariant derivative of gauge* $\mathcal{A}^{\alpha}_{\beta k}$ *itself. Tensor forms* $\mathbb{G}_{km;ln}$ *and* $\Lambda_{kl}$ (*or* $\Lambda^{kl}$) *are named gauge 4-tensor and grand metric* (GM), *respectively. Scalar forms* $\mathbb{N}$, $\mathbb{D}$, $\mathbb{M}$ *and* $\mathbb{G}$ *are named state norm, kinetic scalar, matter scalar and gauge scalar, respectively.*

### *Euler-Lagrange equations on DSV and UGF*

*1. Equations on DSV*:

$$\mathbf{\Lambda}^k \mathfrak{D}_k \mathbf{\Psi} + \left(\frac{1}{2}\mathfrak{D}_k \mathbf{\Lambda}^k + \mathbf{1}\right)\mathbf{\Psi} = 0 ; \qquad (1.10)$$

$$(\mathfrak{D}_k \mathbf{\Phi})\mathbf{\Lambda}^k + \mathbf{\Phi}\left(\frac{1}{2}\mathfrak{D}_k \mathbf{\Lambda}^k - \mathbf{1}\right) = 0 ; \qquad (1.11)$$

here:

$$\mathfrak{D}_k \mathbf{\Lambda}^k \equiv \frac{1}{\sqrt{\Lambda}} \partial_k (\sqrt{\Lambda} \mathbf{\Lambda}^k) + [\mathcal{A}_k , \mathbf{\Lambda}^k] .$$

*2. Equations on UGF*:

$$\mathfrak{D}_l \mathfrak{R}^{kl} = \mathcal{J}^k; \qquad (1.12)$$

$$\mathfrak{D}_l \mathfrak{R}^{kl} \equiv \frac{1}{\sqrt{\Lambda}} \partial_l (\sqrt{\Lambda} \mathfrak{R}^{kl}) + [\mathcal{A}_l, \mathfrak{R}^{kl}] ; \qquad \mathfrak{R}^{kl} \equiv \Lambda^{km}\Lambda^{ln}\mathfrak{R}_{mn} ; \qquad (1.13)$$

$$\mathcal{J}^k \equiv \frac{1}{2}(\mathbf{\Lambda}^k \mathbb{N} + \mathbb{N}\mathbf{\Lambda}^k); \qquad \mathbb{N} \equiv \mathbf{\Phi} \times \mathbf{\Psi}. \qquad (1.14)$$

*Contracted equations*

As the direct outcomes of structure of the SFT objects and EL equations on DSV and UGF, there are the following scalar, vector and tensor equations [1, 2].

1. In dynamics:

$$\mathbb{M} = 0, \quad i.e. \quad \mathbb{D} = -\mathbb{N} ; \qquad (1.15)$$

$$\partial_k (\sqrt{\Lambda} \mathcal{J}^k) = 0 ; \qquad (1.16)$$

$$\frac{1}{\sqrt{\Lambda}} \partial_l (\sqrt{\Lambda} \mathfrak{R}^{kl}) = \mathcal{J}^k. \qquad (1.17)$$

Here:

$$\mathcal{J}^k \equiv Tr\boldsymbol{\mathcal{J}}^k = \Psi^\alpha \Lambda_\alpha^{\beta k} \Phi_\beta \qquad (1.18)$$

$$\mathfrak{R}^{kl} \equiv \Lambda^{km}\Lambda^{ln}\mathfrak{R}_{mn} \qquad (1.19)$$

$$\mathfrak{R}_{kl} \equiv Tr\boldsymbol{\mathfrak{R}}_{kl} = \partial_k \mathcal{A}_{\alpha l}^\alpha - \partial_l \mathcal{A}_{\alpha k}^\alpha . \qquad (1.20)$$

Equations (1.17) are complemented by the connection equations of tensor $\mathfrak{R}_{kl}$ which follow from definition of HCF (1.8):

$$\partial_m \mathfrak{R}_{kl} + \partial_l \mathfrak{R}_{mk} + \partial_k \mathfrak{R}_{lm} = 0 . \qquad (1.21)$$

*Covariance of SFT*

Requirement of compensation for derivatives of matrix $B$ at transformation of equation on $\boldsymbol{\Psi}$ (1.10) leads to *requirement* of the following transformation law for *unified gauge field*:

$$\boldsymbol{\mathcal{A}}' = A^{-1}B(\boldsymbol{\mathcal{A}} + \boldsymbol{\partial})B^{-1}. \qquad (1.22)$$

Consequently, the following transformation laws have been derived in [1] from consideration of dynamic connections given by EL equations on DSV (1.10) and (1.11):

- for *state co-vector* $\boldsymbol{\Phi}$:

$$\boldsymbol{\Phi}' = \boldsymbol{\Phi}B^{-1}; \qquad (1.23)$$

- for covariant derivatives of DSV (1.6):

$$\mathfrak{D}'\boldsymbol{\Psi}' = A^{-1}B\mathfrak{D}\boldsymbol{\Psi}; \qquad \mathfrak{D}'\boldsymbol{\Phi}' = A^{-1}(\mathfrak{D}\boldsymbol{\Phi})B^{-1} ;$$

- for *split metric* matrices $\boldsymbol{\Lambda}$:

$$\boldsymbol{\Lambda}' \equiv AB\boldsymbol{\Lambda}B^{-1}. \qquad (1.24)$$

The supercurrent matrices $\boldsymbol{\mathcal{J}}^k$ then will transform similar to the SM matrices. Further, equations on UGF (1.12) establish connection between *hybrid curvature form* (HCF) (1.8) and $\boldsymbol{\mathcal{J}}^k$. As established in [1], form (1.8) is transformed as a *hybrid tensor* of total valence 4 with two matrices on Roman indices and two matrices on Greek indices:

$$\boldsymbol{\mathfrak{R}}'_{k'l'} = A_{k'}^k A_{l'}^l B\boldsymbol{\mathfrak{R}}_{kl}B^{-1}, \qquad (1.25)$$

if gauge $\mathcal{A}_k$ is transformed according to equation (1.19), and vice versa. So equations (1.12) certify transformation law (1.19) for UGF.

***Comment 1.1.1.*** Principle of *homogeneity* of a fundamental self-contained field theory excludes introduction of any *given in advance* coefficient functions in DSV equations explicit dependent of manifold variables. On the other hand, such terms (associated with derivatives of matrix $B$) arrive at transformation of variables. *Gauge fields* (matrices $\mathcal{A}^{\alpha}_{\beta k}$) are introduced to compensate for this violation. Derivatives of DSV extended in this way become *covariant* to DSV. Thus, *covariance* can be considered an attribute of the *homogeneity* principle.

***Comment 1.1.2.*** Dual State Vector (DSV) of *superdimensional field theory* (SFT) was introduced in paper [1] as a master geometrical object transformed, in accordance with a differential homomorphic concept of *matter function* $\varphi^{\alpha}(\breve{\psi})$, with a matrix $B$ of transformation differentials $d\varphi^{\alpha}$ different from transformation matrix $A$ of UM variables. Generally covariant treat of SFT in [1] based on Lagrangian (1.3) – (1.9), however, has left open a question about connection $B$ to $A$. Principle of covariance (which is, as noticed above, a direct logical consequence of the *homogeneity* demand) itself is not able to specify relations between transformations $B$ and $A$. As it was presumed in paper [1], connection between two matrices has been found in work [2] based on the demand of *transformational invariance* of the SFT differential system.

### 1.2. Reduction of SFT to Invariant Field Theory

*Transformational invariance as a primary demand*

Resolution of issue of connection $B(A)$ has been found in [2] based on the requirement of *transformational invariance* (TI) of form of SFT equations as one of the *irreducibility* demands posed in [2] on the superdimensional theory. In fact, TI is of more fundamental meaning than property of *covariance*. TI covers the commitments of the *covariance* those described in [1] – since introduction of the *unified gauge fields* (UGF) $\mathcal{A}^{\alpha}_{\beta k}$ in the pre-viewed equations for DSV as a measure to compensate the derivatives of the transformation matrix $B$ appears a primer attribute of SFT treat with TI. Moreover, TI covers the demand of homogeneity as well, since introduction or appearance of any explicit functions of UM variables in differential system of SFT is contrary to the TI requirement. In this context, properties of the *homogeneity* and *covariance* can be considered as logical attributes of TI, so they can be replaced by TI in list of requirements to an irreducible field theory in paper [1]; now it will be filled with the following items: *diffeohomomorphism, duality, reality, uniformity, differential irreducibility, transformational invariance, existence of a conservative current, scaling invariance* and *minimax principle*.

*Posing the demand of Transformational Invariance on SFT*

Observing system of the derived covariant EL equations on DSV (1.10), (1.11) and UGF (1.12), one can find that, requirement of transformational invariance of this system is reduced to requirement of constancy of *split metric* matrices $\mathbf{\Lambda}^k$ in result of two commutative transformations: one with matrix $A$ affecting $\mathbf{\Lambda}^k$ on Roman indices, while other with two matrices $B$ on Greek indices:

$$AB\Lambda B^{-1} = \Lambda \quad (1.26)$$

Constancy of SM leads also to constancy of *grand metric* $\Lambda^{kl}$ being structured on SM according to definition (1.9):

$$(\Lambda^{kl})' = \Lambda^{kl}; \quad (1.27)$$

thus, transformational invariance is reduced to the *rotational* one.

*Invariant reduction of SM* [2]

Implementation of TI has been produced in [2] in terms of *infinitesimal* transformation:

$$A_l^k = 1 + a_l^k; \quad (1.28)$$

restriction (1.27) then leads to the following conditions on infinitesimal matrix $a_l^k$:

$$a_k^n \Lambda_{ln} + a_l^n \Lambda_{kn} = 0, \quad (1.29)$$

or

$$a_{kl} = -a_{kl}; \qquad a_{kl} \equiv a_k^n \Lambda_{ln} \quad (1.30)$$

Equations (1.26) then lead to solution for the related matrix $B = 1 + b$:

$$b = \frac{1}{4}\sigma^{kl} a_{kl}; \qquad \sigma^{kl} \equiv \frac{1}{2}(\Lambda^k \Lambda^l - \Lambda^l \Lambda^k) \quad (1.31)$$

under a condition that SM matrices $\mathbf{\Lambda}^k$ satisfy the following requirements:

$$\mathbf{\Lambda}^k \mathbf{\Lambda}^l + \mathbf{\Lambda}^l \mathbf{\Lambda}^k = 2\Lambda^{kl} \cdot \mathbf{1}. \quad (1.32)$$

In *orthogonal reduction* conditions (1.32) can be written as follows:

$$\mathbf{\Lambda}^k \mathbf{\Lambda}^l + \mathbf{\Lambda}^l \mathbf{\Lambda}^k = \begin{cases} 0; & l \neq k \\ 2\Lambda^{kk} \cdot \mathbf{1}; & l = k. \end{cases} \quad (1.33)$$

$$(\mathbf{\Lambda}^k)^2 \equiv \Lambda^{kk} \cdot \mathbf{1}; \quad (1.34)$$

$$\boldsymbol{\sigma}^{kl} = \mathbf{\Lambda}^k \mathbf{\Lambda}^l; \quad l \neq k. \quad (1.35)$$

Terms $\Lambda^{kk}$ on definition are squares of matrices $\mathbf{\Lambda}^k$ proportional to unit matrix; they play role of metric tensor (*grand metric*, GM) reduced to a diagonal form.

*Split Metric matrices as proportions of Cartan's affinors*

Considering algebraic properties of SM matrices, one can introduce normalized affinors $\mathbf{A}^k$:

$$\mathbf{\Lambda}^k \rightarrow \frac{\mathbf{\Lambda}^k}{\sqrt{|\Lambda^{kk}|}} \equiv \mathbf{A}^k ; \tag{1.36}$$

they will satisfy the normalized equations:

$$\mathbf{A}^k \mathbf{A}^l + \mathbf{A}^l \mathbf{A}^k = \begin{cases} 0; & l \neq k \\ a^k \cdot \mathbf{1}; & l = k; \quad a^k = \pm 1 \end{cases} \tag{1.37}$$

Affinors $\mathbf{A}^k$ correspond to matrices $A^k$ of E. Cartan *Theory of Spinors* [3], with two distinctions: 1) they may vary in space of a manifold; 2) affinors $A^k$ are defined for all $a^k = 1$, since, in his mathematical foundation and treat of spinors, metric of space is not determined by affinors $A^k$.

## *Clifford algebra of orthogonal SM*

Orthogonal conditioning of SM (1.33) directly brings *Clifford algebra* of a collection of matrices obtained as various power multi-products of SM matrices $\mathbf{\Lambda}^k$, with total number of independent irreducible products $2^N$ [2]. Isomorphic identity of collection of such matrices to arbitrary matrix of rank $\mu$ requires an equity: $2^N = \mu^2$, then $\mu = 2^{N/2}$; so $N$ must be *even*. In this way, posing TI on SFT leads to finding *connection of dimensions N and $\mu$* [2, 3, 4]. We call this connection *structural isomorphism* of SSFT.

### *Comment 1.2.1. On reduction of **transformational** invariance principle to the **rotational** one*

Reduction of the *transformational invariance* (TI) requirement to the *rotational* one may look like admission of restrictions or limitations to the *general relativity* principle (GRP) in frame of *irreducible* superdimensional field theory concept under investigation. One might blame for this restriction the fact that metric tensor of the theory is structured on SM according to definition (1.9). It should be underlined, however, that, tensor $\Lambda^{kl}$ *always exists* regardless to a role that this tensor is committed to play in structure and dynamics of the theory, − once there is a *hybrid tensor* (*split metric* matrices) $\mathbf{\Lambda}^k$ as *coefficient functions* in equations for dual *state vector field* (DSV). On the other hand, one may raise a question, should not GRP be associated only with an *irreducible* class of transformations of variables of a manifold? In the context of the demand of *irreducibility*, resorting to TI principle at formulation of an irreducible field theory is totally equivalent to a direct resorting to the *rotational invariance* (RI) of the *split metric* $\mathbf{\Lambda}^k$. So extension of RI principle to more general transformational invariance (besides such a trivial one as *shift*) is not possible; after all, it is not required and cannot be logically motivated by a substantial reason.

### *Comment 1.2.2. Orthogonal Split Metric as genetic precursor of notion of an orthogonal frame in SSFT*

In our treat of SFT and related geometry based on the dynamical principles as the primary ones, *orthogonality* (i.e. diagonal form of *grand metric* tensor (1.9)) arrives as an attribute of irreducible *orthogonal* reduction of relations between affinors $\mathbf{\Lambda}^k$ in accordance with E. Cartan's *theory of spinors*. Setting matrices $\mathbf{\Lambda}^k$ as an internally *orthogonal* collection (definition!) according to equations (1.33) can be considered as an *irreducible utilization of the TI principle* in the differential dynamical system of the superdimensional FT. GM then becomes diagonal, thus leading to reduction of *interval form* to a *quadratic* one, and to characterization of initially unspecified coordinate frame as an *orthogonal* one. Also, it should be noted that, orthogonal formulation of relations between affinors $\mathbf{\Lambda}^k$ is objectively inquired, since it allows one to immediately and simply recognize *Clifford algebra* of matrices $\mathbf{\Lambda}^k$.

Thus, principle of *transformational invariance* imposed to SFT leads to explication of the geometrical nature of matrices $\mathbf{\Lambda}^k$ as proportions of *Cartan's spin affinors*.

### 1.3. Finite rotations

Finite rotations can be found by integration of the infinitesimal equations based on the *group principle* of the rotations; that suggests introduction of rotation parameters $\xi$ in a way that

$$A(\xi + \Delta\xi) = A(\Delta\xi)A(\xi).$$

If matrix of infinitesimal rotation is $1 + a$, then differential of matrix $A$ is equal to $aA$; similar relation is valid for transformation of DSV with matrix $B$. As it has been clarified in [2], group principle can be utilized at the following parameterization of matrix $a$ in an orthogonal frame:

$$a_l^k = \frac{\sqrt{|\Lambda_{kk}\Lambda_{ll}|}}{\Lambda_{kk}} d\varphi_{kl}; \qquad a_k^l = -\frac{\sqrt{|\Lambda_{kk}\Lambda_{ll}|}}{\Lambda_{ll}} d\varphi_{kl} \qquad (1.38)$$

where $\varphi_{kl}$ ("rotation angle") is *group parameter* for rotation in plane $(k, l)$. Then, according to relations (1.36),

$$b = \frac{1}{4}\mathbf{A}^k\mathbf{A}^l d\varphi_{kl}. \qquad (1.39)$$

Transformations of frame and DSV at finite rotations in a single plane $(k, l)$ at point of UM have been found in paper [2].

*Rotation of a UM plane*

$$A^{(kl)} = \begin{pmatrix} \cos\varphi_{kl}; & \Lambda_k^l \sin\varphi_{kl} \\ -\Lambda_l^k \sin\varphi_{kl}; & \cos\varphi_{kl} \end{pmatrix}; \qquad \Lambda^{kk} \cdot \Lambda^{ll} > 0 \qquad (1.40)$$

$$A^{(kl)} = \begin{pmatrix} ch\varphi_{kl}; & \Lambda_k^l sh\varphi_{kl} \\ \Lambda_l^k sh\varphi_{kl}; & ch\varphi_{kl} \end{pmatrix}; \qquad \Lambda^{kk} \cdot \Lambda^{ll} < 0 \qquad (1.41)$$

here:

$$\Lambda_l^k \equiv \sqrt{\left|\frac{\Lambda^{kk}}{\Lambda^{ll}}\right|}. \tag{1.42}$$

*Rotation of DSV*:

$$B^{(kl)} = \cos\frac{\varphi_{kl}}{2} + \mathbf{A}^k \mathbf{A}^l \sin\frac{\varphi_{kl}}{2} ; \qquad \Lambda_{kk}\Lambda_{ll} > 0 \tag{1.43}$$

$$B^{(kl)} = ch\frac{\varphi_{kl}}{2} + \mathbf{A}^k \mathbf{A}^l sh\frac{\varphi_{kl}}{2} ; \qquad \Lambda_{kk}\Lambda_{ll} < 0. \tag{1.44}$$

As presumed in paper [2] (see comment 1.2.1. above), posing the TI requirement on differential system of SFT leads to *rotational* invariance of SM and to explication of SM and DSV as *spin-affinors* of E. Cartan *theory of spinors* [3 – 5] and *spin-vector field* as an extended analog of Dirac's *spin function* of a relativistic electron. In case of dimensionality $N > 4$, DSV field then can be characterized as a *superdimensional dual Fermi-Dirac field*, briefly *superspinor*.

Transformation of *unified gauge field* (UGF) at rotation in a plane $(k, l)$ has been shown in [2] applying general law (1.22):

$$\mathcal{A}' = (A^{(kl)})^{-1} B^{(kl)} (\mathcal{A} + \partial)(B^{(kl)})^{-1} \tag{1.45}$$

Transformations of UGF at finite rotations have been illustrated assuming that, rotation of frame of *unified manifold* (UM) is characterized by parameter $\varphi_{kl}$ constant in space, then:

$$\partial (B^{(kl)})^{-1} = -(\partial \boldsymbol{\sigma}^{(kl)}) \cdot \begin{cases} \sin\dfrac{\varphi_{kl}}{2} ; & \Lambda_{kk}\Lambda_{ll} > 0 \\ sh\dfrac{\varphi_{kl}}{2} ; & \Lambda_{kk}\Lambda_{ll} < 0. \end{cases} \tag{1.46}$$

It is worth to note that, matrix $B$ is still inhomogeneous in UM space even at constant (homogeneous in space of UM) rotation parameters $\varphi$ due to that Split Metric matrices $\boldsymbol{\Lambda}$ vary in space being connected to DSV and UGF. Terms with derivatives of rotation parameters $\varphi_{kl}$ arrive if there is a reason or necessity to consider their variation in space. Regardless of possibility of such extension, transformation of UGF (1.41) as being given by binary products of elements of matrix $B$ should be expressed in result in terms of unambiguous analytical functions of rotation parameter $\varphi$ of UM ($sin\varphi, cos\varphi; sh\varphi, ch\varphi$), in contrary to transformation of DSV given by the first power of matrix $B$ (1.43), (1.44). So UGF can be envisioned corresponding to the *boson* class of QFT objects i.e. "elementary particles" of an *integer spin*. By the way, boson type of objects can also be associated with *clusters* i.e. *hybrid tensors* structured on an even number of DSV components in product.

*General equations of rotations*

General infinitesimal rotation of frame in UM space is given by $\frac{N(N-1)}{2}$ parameters $\varphi_{kl}$. One can derive differential equation considering a *path line* of transformation $\varphi_{kl}(\xi)$ which is characterized by parametric derivatives $\omega_{kl}$:

$$d\varphi_{kl} = \omega_{kl}(\xi)d\xi; \quad \omega_{kl} = -\omega_{lk}. \tag{1.47}$$

Matrix $a_l^k$ of infinitesimal rotation now can be represented through "velocities" $\omega_{kl}$:

$$a_l^k = \Omega_l^k d\xi; \tag{1.48}$$

$$\Omega_l^k \equiv \frac{\sqrt{|\Lambda_{kk}\Lambda_{ll}|}}{\Lambda_{kk}}\omega_{kl}; \quad \Omega_k^l = -\frac{\sqrt{|\Lambda_{kk}\Lambda_{ll}|}}{\Lambda_{ll}}\omega_{kl}; \quad \Omega_k^k = 0 \tag{1.49}$$

(no summation on $k, l$), so we obtain differential equation for matrix $A \equiv A_l^k$ (summation on $n$):

$$\frac{dA_l^k}{d\xi} = \Omega_n^k(\xi)A_l^n. \tag{1.50}$$

A corresponding equation can be derived for matrix $B$ [2]:

$$\frac{dB}{d\xi} = \frac{1}{4}\omega_{kl}(\xi)\mathbf{A}^k\mathbf{A}^l B \tag{1.51}$$

(summation on $k$ and $l$).

### 1.4. Lagrangian and metric equations of invariant SFT

In the SFT concept matrices $\mathbf{\Lambda}^k$ as coefficient functions in equations on DSV are treated as subjects of independent variations in the extreme action; in this way they arrive connected to DSV and UGF by the correspondent Euler-Lagrange equations. Posing principle of transformation invariance of SFT differential system does not allow one to consider these $N$ matrices completely as subjects of independent variations – since they are subordinate of demands (1.33). These internal correlations between $\mathbf{\Lambda}^k$ have been taken into account in [2] applying techniques of *Lagrange multipliers* of EAP when it is constrained with connection between objects of variations:

$$\mathbb{L} \Longrightarrow \mathbb{L} + \mathbb{L}_{TI}; \tag{1.52}$$

$$\mathbb{L}_{TI} \equiv \frac{1}{2}Tr(\mathbf{M}_{kl}\mathbf{C}^{kl}); \tag{1.53}$$

$$\mathbf{C}^{kl} \equiv \mathbf{\Lambda}^k\mathbf{\Lambda}^l + \mathbf{\Lambda}^l\mathbf{\Lambda}^k - 2\Lambda^{kl}\cdot\mathbf{1}. \tag{1.54}$$

Form $\mathbb{L}_{TI}$ is scalar Lagrangian of the *transformational invariance*; we call matrices $\mathbf{C}^{kl}$ *Cartan's constraints form*; matrices $\mathbf{M}_{kl}$ are considered as the additional independent objects of the *extreme action* playing an intermediate role in the differential system as Lagrange multipliers. Scalar $\mathbb{L}_{TI}$ does not include DSV and UGF, so EL equations for these objects do not change, while equations on SM have been significantly modified compared with equations derived in [1].

*EL equations on Lagrange multipliers* $\mathbf{M}_{kl}$ :

$$\frac{\partial \mathcal{L}}{\partial \mathbf{M}_{kl}} = 0$$

simply manifest in *spin-equations* (1.33).

*EL equations on SM*

Taking into account that *in dynamics* $\mathbb{M} = 0$ and $\mathbb{L}_{TI} = 0$, EL equations on SM:

$$\frac{\partial \mathcal{L}}{\partial \mathbf{\Lambda}^k} = 0$$

can be written in the following view [2]:

$$(\mathbb{G}_{kl} - \mathbb{G}\Lambda_{kl})\mathbf{\Lambda}^l + \mathbf{M}_{kl}\mathbf{\Lambda}^l + \mathbf{\Lambda}^l\mathbf{M}_{kl} - 2\mathbf{\Lambda}^l Tr\mathbf{M}_{kl} = -\mathfrak{D}_k . \qquad (1.55)$$

***Comment 1.4.1.*** *Derivability of non-linear algebraic equations for Grand Metric tensor*

As discussed in [2] system of equations (1.55) and (1.33) (algebraic relative *split metric* matrices $\mathbf{\Lambda}^k$) can be solved relative $\mathbf{\Lambda}^k$ as functions of DSV, UGF and GM $\Lambda^{kl}$. Using then definition of GM (1.9), one can obtain algebraic equations for GM tensor as function of tensors structured on DSV, UGF and their derivatives.

*Constraint of regional parameterization of rotations*

Constraint of explicit analytical representation of finite rotations in Riemannian space of UM suggests parameterization of rotations with parameters $\varphi_{kl}$ constant in space. Formally, such parameterization can be interpreted as related to finite rotation in *tangent* (pseudo)Euclidian space at a point of UM [7]; however, such methodological trick is artificial yet difficult to be supported as itself by a direct physical reasoning. One needs to find out a way to connect parameters of a finite transformation by a functional algorithm over a region of UM space, either particular or general. By the way, there is a consistent yet fundamental way to bring a direct dynamical sense for regional parameterization with angle parameters constant in Riemannian space of *unified manifold* (UM). It is based on consideration of the *geodesic lines* in UM; they embody the *Riemannian coordinates* (RC) [7, 8]. Transformations in terms of RC are equivalent to linear transformations in a (pseudo)Euclidian space – though a flat geometry space has no real physical meaning.

## 2. Geodesics in Unified Manifold of SSFT

*Lines in UM*

A *line* in UM is given by a system of parametric equations for *coordinates* $\check{\psi}^k$:

$$\check{\psi}^k \longrightarrow \check{\psi}^k(\tau); \quad k = 1, 2, \ldots N, \tag{2.1}$$

where $\tau$ is a *canonical parameter*, a continuously variable numerical argument. *Line direction* at a point $\check{\psi} \equiv \{\check{\psi}^k\}$ is determined by derivatives $\check{\psi}^k(\tau)$ on $\tau$, the *tangent vector* (TV):

$$\xi^k(\tau) \equiv \frac{d\check{\psi}^k(\tau)}{d\tau}. \tag{2.2}$$

One can consider a continuous set of lines and correspondent set of TV i.e. TV *field* $\xi^k(\check{\psi})$ as function of UM variables. TV field can be subordinate of a differential law associated with covariant derivatives of TV, tensor $\mathcal{D}_l \xi^k$:

$$\mathcal{D}_l \xi^k \equiv \partial_l \xi^k + \Gamma^k_{ml} \xi^m; \tag{2.3}$$

$$\Gamma^k_{lm} \equiv \frac{1}{2} \Lambda^{kn}(\partial_l \Lambda_{mn} + \partial_m \Lambda_{ln} - \partial_n \Lambda_{lm}). \tag{2.4}$$

*Absolute differential of a tangent vector*

*Absolute differential* (AD) of a vector field $\xi^k(\check{\psi})$ is an infinitesimal form:

$$\check{d}\xi^k \equiv \mathcal{D}_l \xi^k d\check{\psi}^l \tag{2.5}$$

which can be applied to determine change of TV at infinitesimal displacement in a direction $\eta^l$: if

$$d\check{\psi}^l \Rightarrow d_\eta \check{\psi}^l = \eta^l d\tau, \tag{2.6}$$

then

$$\check{d}\xi^k \Rightarrow \check{d}_\eta \xi^k = \eta^l \mathcal{D}_l \xi^k d\tau = (\eta^l \partial_l \xi^k + \Gamma^k_{ml} \xi^m \eta^l) d\tau \tag{2.7}$$

*AD in direction of the line*

AD in direction of tangent vector $\boldsymbol{\xi}$ itself (*tangent AD*) is of special interest:

$$\mathcal{D}\xi^k \Rightarrow \check{d}_\xi \xi^k = \xi^l \mathcal{D}_l \xi^k d\tau = (\xi^l \partial_l \xi^k + \Gamma^k_{ml} \xi^m \xi^l) d\tau; \tag{2.8}$$

taking into account an identity:

$$\xi^l \partial_l \xi^k = \frac{d\check{\psi}^l}{d\tau} \frac{\partial \xi^k}{\partial \check{\psi}^l} \equiv \frac{d\xi^k}{d\tau} \tag{2.9}$$

One finds the following expression for tangent AD:

$$\check{d}_\xi \xi^k = \left( \frac{d\xi^k}{d\tau} + \Gamma^k_{ml} \xi^m \xi^l \right) d\tau \tag{2.10}$$

*The geodesic lines*

Our definition of a *geodesic line*: tangent AD of the line is equal zero:

$$\frac{d\xi^k}{d\tau} + \Gamma^k_{ml} \xi^m \xi^l = 0 . \tag{2.11}$$

Since object $\Gamma^k_{lm}$ is considered a function of variables $\check{\psi}^k$, we have to return in equations (2.11) to definition (2.2) of tangent vector $\xi^k$, then we obtain *equations of geodesic lines* as follows:

$$\frac{d^2 \check{\psi}^k}{d\tau^2} + \Gamma^k_{lm} \frac{d\check{\psi}^l}{d\tau} \frac{d\check{\psi}^m}{d\tau} = 0 . \tag{2.12}$$

### 3. Riemannian coordinates

Riemannian coordinates are introduced in the following way [7]. Take a point $M_0$ in a manifold with matched connection $\Gamma^k_{lm}(\check{\psi})$ and drive geodesic lines crossing the point in all directions; every geodesic line is characterized by the "start" tangent vector:

$$\xi^k_0 = \left( \frac{d\check{\psi}^k}{d\tau} \right)_0 \tag{3.1}$$

and satisfies equations (2.12). A point $M$ on an arbitrary geodesic line can be characterized by $N$ numbers

$$\xi^k_0 \tau \equiv y^k ; \tag{3.2}$$

these numbers are called the *Riemannian coordinates*. It is important that, at an arbitrary transformation of coordinates $\check{\psi}^k \to \check{\psi}^{k\prime}$, RC are transformed as a contravariant vector:

$$y^{k\prime} = \xi^{k\prime}_0 \tau = (A^{k\prime}_k)_0 \xi^k_0 \tau = (A^{k\prime}_k)_0 y^k . \tag{3.3}$$

Thus, transformation between two different RC frames is linear, i.e. the correspondent matrix $A_{RC}$ is constant in space of UM:

$$A_{(RC)} = const. \tag{3.4}$$

Equations (3.2) can be interpreted as parametric equations of geodesic lines (that cross point 0) in the Riemannian coordinates; these equations are linear in canonical parameter $\tau$. On the other hand, equations of geodesic lines can also be written in the Riemannian coordinates in general form as follows:

$$\frac{d^2 y^k}{d\tau^2} = -\Gamma^k_{lm} \frac{dy^l}{d\tau} \frac{dy^m}{d\tau} . \qquad (3.5)$$

But, according to the parametric equations for RC (3.2),

$$\frac{dy^k}{d\tau} = \xi^k = const ; \qquad (3.6)$$

from here we conclude:

$$\frac{d^2 y^k}{d\tau^2} = 0 , \qquad (3.7)$$

hence, as it follows from the equation for geodesic lines (3.5):

$$\Gamma^k_{lm} \frac{dy^l}{d\tau} \frac{dy^m}{d\tau} = \Gamma^k_{lm} \xi^l \xi^m = 0. \qquad (3.8)$$

Multiplying this equation by $\tau^2$, we obtain:

$$\Gamma^k_{lm} y^l y^m = 0 . \qquad (3.9)$$

The last equations can be considered as *general definition* of the *Riemannian coordinates.* Note that, the Riemannian coordinates are also geodesic lines relative to a chosen initial point $M_0$. This means that $\Gamma^k_{lm}$ in RC is equal zero at this point. To prove this, note that, being an object of the unified manifold, matched connection in the Riemannian coordinates (RC) can be considered a function of variables $y^k$: $\Gamma^k_{lm} \to \Gamma^k_{lm}(y)$, so at the initial point, we have $\Gamma^k_{lm}(0)$. On the other hand, equation (3.9) is valid at an arbitrary point, including $y^k = 0$. Since the direction of the tangent vector $\xi^k$ is arbitrary, we have to accept: $\Gamma^k_{lm}(0) = 0$.

## 4. SSFT in Riemannian coordinates

### 4.1. Differential Law in terms of RC

The EL differential system of SFT is invariant relative of arbitrary transformation of the manifold variables, so all the objects can be immediately considered as functions of RC, but the system has to be complemented by $N$ conditions (3.9).

### 4.2. Rotations and Transformations of objects in RC

According to equations (3.3), matrix $A$ of transformation in RC being determined by transformation (rotation) of coordinates at "start point" is constant in a region (or overall) of UM space. So transformations (rotations) in terms of an RC frame are *linear* with respect to Roman indices (but not so with respect to the Greek ones).

### 4.3. Transformation of objects in an orthogonal RC frame

At *orthogonal* definition of Split Metric matrices $\mathbf{\Lambda}^k$ according to equations (1.30), (1.31), metric tensor has only diagonal non-zero components $\Lambda^{kk}$. Consequently, rotations in a plane $(y^k, y^l)$ are characterized by parameters $\omega_{kl}$ (or $\varphi^{kl}$) and $\Lambda_l^k \equiv \sqrt{|\Lambda^{kk}/\Lambda^{ll}|}$ of matrix $A$ according to expressions (1.38). Since matrix $A$ in terms of RC is constant in space of UM, expressions (), () are immediately expandable over whole UM space with constant both $\omega_{kl}$ (or $\varphi_{kl}$) and $\Lambda_l^k$:

$$\omega_{kl} = const; \quad \varphi_{kl} = const; \qquad (4.1)$$

$$\Lambda_l^k \equiv \sqrt{|\Lambda^{kk}\Lambda_{ll}|} = const. \qquad (4.2)$$

We call property (4.2) *conformal invariance of GM*. It is an attribute of an *orthogonal* RC (ORC) frame. Underline that, notion and utilization of ORC frame is associated with structuring of SM matrices according to equations for *affinors* $\mathbf{\Lambda}^k$ (1.33), (1.34).

## 5. An orthonormal RC frame

### 5.1. Utilization of conformal invariance

*Conformal invariance* of GM in an ORC frame can be explicated by the following representation of GM:

$$\Lambda_{kk} = \lambda_k \sqrt[N]{\Lambda}; \qquad (5.1)$$

then

$$\lambda_k = const. \qquad (5.2)$$

Property (5.2) can be simply proved, taking into account that, on definition:

$$\Lambda = \left|\prod_{k=1}^{N} \Lambda_{kk}\right|, \qquad (5.3)$$

so

$$\prod_{l}^{N} |\lambda_l| = 1; \qquad (5.4)$$

Picking any particular $\lambda_k$, we can rewrite (5.4) as follows:

$$|\lambda_k|^N \prod_l^N \left|\frac{\lambda_l}{\lambda_k}\right| = 1; \qquad (5.5)$$

since, according to definition of parameters $\lambda_k$, all their ratios are constant in space, then so is $(\lambda_k)^N$, hence, property (5.2) is proved.

## 5.2. Normalization of an orthogonal RC metric

Property of *conformal invariance* of metric (4.2) in orthogonal RC does prompt the following normalization of an orthogonal RC metric. Let there is an "initial" collection of parameters $\lambda^k$. Considering EL equations on DSV (), let us produce the following scaling transformation of variables $\check{\psi}^k$:

$$\check{\psi}^k \to \check{\psi}'^k = \sqrt{|\lambda^k|}\check{\psi}^k, \qquad (5.6)$$

and consider products $\sqrt{|\lambda^k|}\mathbf{\Lambda}^k$ and $\mathcal{A}_k/\sqrt{|\lambda^k|}$ as new, renormalized SM and UGF matrices, respectively. Then, according to definition of parameters $\lambda_k$ (note that, $\lambda^k = \frac{1}{\lambda_k}$), we obtain the following normalization for $\Lambda_{kk}$:

$$|\Lambda_{kk}| = \sqrt[N]{\Lambda} \quad \to \quad \Lambda_{kk} = \lambda^k \sqrt[N]{\Lambda} = \pm\sqrt[N]{\Lambda} \qquad (5.7)$$

i.e. our new parameters $\lambda^k$ simply are *metric signature*:

$$\lambda^k \Rightarrow \pm 1. \qquad (5.8)$$

*Metric signature* $\lambda^k$ is invariant of rotations, together with $\mathbf{\Lambda}^k$ and $\Lambda$. Such normalization of matrices $\mathbf{\Lambda}^k$ always can be accepted; once so, equations (1.55) and (1.33) can be used to derive equations for *determinant* of metric tensor as discussed in section 1.3.; signature $\lambda^k$ there will play role of discrete parameters. Choice of one or another collection of $\lambda^k$ (together with selections of dimensionality $N$) is presupposed to be motivated by considerations of consistence, irreducibility and resolvability of the SSFT dynamic equations in all *real* solutions.

## 6. Transformations in orthonormal RC

*Rotation of a UM plane*

Rotation of an orthonormal RC frame in a plane $(k, l)$ is given by a standard matrix of rotation in (pseudo)Euclidian space:

$$A^{(kl)} = \begin{pmatrix} \cos\varphi_{kl}; & \sin\varphi_{kl} \\ -\sin\varphi_{kl}; & \cos\varphi_{kl} \end{pmatrix}; \quad \lambda^k \cdot \lambda^l > 0 \qquad (6.1)$$

$$A^{(kl)} = \begin{pmatrix} ch\varphi_{kl}; & sh\varphi_{kl} \\ sh\varphi_{kl}; & ch\varphi_{kl} \end{pmatrix}; \qquad \lambda^k \cdot \lambda^l < 0 \tag{6.2}$$

*Rotation of DSV*

The correspondent rotation matrix $B$ of DSV has same view as (), ():

$$B^{(kl)} = \cos\frac{\varphi_{kl}}{2} + \mathbf{A}^k \mathbf{A}^l \sin\frac{\varphi_{kl}}{2}; \qquad \lambda_k \lambda_l > 0 \tag{6.3}$$

$$B^{(kl)} = ch\frac{\varphi_{kl}}{2} + \mathbf{A}^k \mathbf{A}^l sh\frac{\varphi_{kl}}{2}; \qquad \lambda_k \lambda_l < 0. \tag{6.4}$$

but now

$$\mathbf{A}^k \equiv \Lambda^k \sqrt[2N]{\Lambda}.$$

*Transformation of UGF*

Transformation of gauge matrices $\mathcal{A}_k$ is given by same formulas (1.45), (1.46).

***Comment 6.1.*** *Texture of Riemannian geometry in an orthogonal normalized RC frame*

*Riemannian geometry* (RG) in terms of an *orthogonal normalized frame* of coordinates has been treated by E. Cartan [6]. *Riemannian coordinates* (RC) can be introduced regardless of a consideration of connection of metric tensor to objects of a physical field theory [7, 8]. Treat of RG in terms of an orthogonal normalized RC frame, as profiled in this paper, reduces texture of RG to *signature* of metric tensor and evolution of its *determinant* in space of a manifold.

## 7. Matched Connection in an orthonormal RC frame

We now can specify structure and transformation property of Matched Connection (MC) (2.4).

1. MC in the RC terms is *tensor*, since matrix $A$ does not change in the UM space.

Considered as an object in the unified manifold i.e. functions of manifold variables, RC themselves are transformed as a contra-variant vector with the same constant matrix in all area of their definition, as pointed above by equations (6.1), (6.2).

2. MC (2.4) being written in terms of an orthogonal frame of coordinates $(i.e. \Lambda^{kl} = 0, k \neq l)$ is reduced to the following view:

$$\Gamma^k_{lm} = \frac{1}{2}\Lambda^{kk}\bigl(\Delta^k_m \partial_l \Lambda_{mm} + \Delta^k_l \partial_m \Lambda_{ll} - \partial_k \Lambda_{lm}\bigr). \tag{7.1}$$

3. In an RC frame, taking into account metric normalization (5.3), MC acquires the following form (no summation on $k, l$):

$$\Gamma^k_{lm} = \frac{1}{2N\Lambda}\left(\Delta^k_m \partial_l + \Delta^k_l \partial_m - \Delta_{lm}\frac{\lambda_l + \lambda_m}{2\lambda_k}\partial_k\right)\Lambda. \tag{7.2}$$

4. Equations of Riemannian coordinates (3.9) are then specified to the following view (no summation on $k$):

$$\sum_l y^l \left(y^k \partial_l - y^l \frac{\lambda_l}{2\lambda_k}\partial_k\right)\Lambda = 0. \tag{7.3}$$

## 8. Resume

*Summary*

Purpose of this work was to explore possibilities of an irreducible formulation of the *theory of transformations* of the *superspinorial field theory* (SSFT) presented in work [2]. Main results of the exploration are in the following.

1. Method of *Riemannian coordinates* (RC) based on resorting to *geodesics* allows one to reduce the *global* texture of transformations in (pseudo)Riemannian space to *linear transformations* as *rotations* of coordinate frame of the *unified manifold*, matrix of which is *constant in space*.

2. Form of SSFT equations derived in [2] does not change at transition to RC but is complemented with *RC equations* (6.3).

3. In terms of RC, the irreducible *orthogonal* definition *of split metric* (SM) matrices $\mathbf{\Lambda}^k$ in accordance with *Cartan's affinors of Theory of Spinors* makes the whole *metric theory of SSFT* formulated as theory of metric *determinant* and *signature.*

4. In contrary to constancy of matrix $A$ of rotation of RC frame, matrix $B$ of rotation of the dual *state vector field, superspinor* (DSV), being dependent of matrices $\mathbf{\Lambda}^k$, varies with $\mathbf{\Lambda}^k$ in UM space as determined by equations (1.55) and (1.33).

Thus, with transition to RC, *spinorial reduction* of SFT to SSFT produced preliminary in [2] acquires a final irreducible formulation.

*Outlook*

Next step of studying SSFT should be search for solution of algebraic *metric equations* (1.55), (1.33) relative *split metric* matrices $\mathbf{\Lambda}^k$ as discussed in section 1.4. Having this in hands, one could explore two critical aspects of the theory: 1) asymptotic solutions of non-linear equations for DSV and UGF as possibly associated with notion of *elementary particles* (EP); 2) asymptotic behavior of *grand metric* tensor $\Lambda^{kl}$ far of *matter as neutral clusters of EPs*, to check possibility of profiling *gravitation* as a *macro-phenomenon* in the projective 4 dimensions space-time manifold as intelligible world.